\documentclass[a4paper,11pt]{article}

\usepackage[utf8]{inputenc}

\usepackage[margin=1in]{geometry}      
\usepackage{verbatim}
\usepackage{graphicx}
\usepackage{amssymb}
\usepackage[bookmarks=true]{hyperref}
\hypersetup{
	colorlinks=true,
	linkcolor=orange,
	citecolor=blue,
	filecolor=magenta,
	urlcolor=magenta
 ,linktocpage=true
,pdfproducer=medialab
,pdfa=true
}
\usepackage{amsmath}
\usepackage{amsthm}
\usepackage{physics}
\usepackage{extarrows}
\usepackage{amsfonts}
\usepackage{mathrsfs}
\usepackage{authblk}
\usepackage{cancel}
\usepackage{subfigure}
\usepackage[dvipsnames]{xcolor}
\usepackage{comment}
\usepackage{multirow}
\usepackage{cite} 
\usepackage{bigints}

\newcommand{\beq}{\begin{equation}}
\newcommand{\eeq}{\end{equation}}
\newcommand{\bea}{\begin{eqnarray}}
\newcommand{\eea}{\end{eqnarray}}

\newcommand{\ord}[1]{\mathcal{O}({#1})} 

\newcommand{\fref}[1]{Figure~\ref{#1}} 
\newcommand{\eref}[1]{Eq.\,(\ref{#1})}

\newcommand{\sref}[1]{Section~\ref{#1}}
\newcommand{\tref}[1]{Table~\ref{#1}}

\begin{document}

\mbox{}\vspace{1cm}

\begin{center}
    {\LARGE \bf \boldmath Testing axion couplings to leptons \\[1mm] in $Z$ decays at future $e^+e^-$ colliders} \\[1.2cm]	{\large
	Lorenzo Calibbi, 
    Zijie Huang,  
	  Shaoyang Qin, \\[1mm] 
	Yiming Yang,\footnote{Corresponding author, {\it e-mail}\,: \href{mailto:yang_yiming@mail.nankai.edu.cn}{yang\_yiming@mail.nankai.edu.cn}}
    and Xiaoyue Yin 
	}\\[0.5cm]
 	{
       \it   School of Physics, Nankai University, Tianjin 300071, China  \\[5pt]}
\end{center}
\vspace*{0.8cm}

\begin{abstract}
\noindent
\noindent We study the possibility of probing the existence of a light, invisible, axion-like particle (ALP) $a$ in leptonic decays of the $Z$ boson at the proposed high-energy $e^+e^-$ colliders, CEPC and FCC-ee.
Both projects plan to run at the $Z$ pole, collecting $10^{12}-10^{13}$ visible $Z$ decays. We show that, searching for the emission of an invisible ALP from leptons in leptonic $Z$ decays, this enormous statistics could allow to constrain the ALP couplings to leptons at an unprecedented level for laboratory experiments. 
In particular, within a Monte Carlo simulation framework, we estimate that CEPC/FCC-ee can be sensitive to the coupling of an invisible ALP to muons up to $f_a/C^A_{\mu\mu} \approx 1$~TeV\,\,\,---\,\,\,where $f_a$ is the ALP decay constant\,\,\,---\,\,\,corresponding to \mbox{${\rm BR}(Z \to \mu^+\mu^-\,a) \approx 3\times 10^{-11}$}.  
\end{abstract}
\vspace*{0.5cm}


\thispagestyle{empty}

\newpage

\setcounter{tocdepth}{1}
\tableofcontents


\setcounter{footnote}{0}

\section{Introduction}

Despite the undeniable success of the Standard Model (SM) and the absence at present of conclusive evidence for new physics (NP) at the Large Hadron Collider (LHC) and low-energy experiments, a number of experimental observations and open problems call for its extension\,\,\,---\,\,\,among  others: dark matter, neutrino masses, matter-antimatter asymmetry, the flavour puzzle.

Interestingly, a wide class of models that have been proposed to address these questions predicts the existence of so-called axion-like particles (ALPs), light weakly-coupled pseudoscalar fields that are the pseudo-Nambu-Goldstone bosons (PNGBs) of spontaneously broken global symmetries. The prototypical example is the axion, which arises from the breaking of the Peccei-Quinn $U(1)$ symmetry that was introduced to solve the strong CP problem~\cite{Peccei:1977hh,Wilczek:1977pj,Weinberg:1977ma}. 
Similarly, other well-motivated ALPs are light remnants of high-energy global symmetries associated with further outstanding problems of the SM, such as the lepton number within neutrino mass models~\cite{Chikashige:1980ui,Gelmini:1980re,Georgi:1981pg}, or flavour symmetries~\cite{Davidson:1981zd,Wilczek:1982rv,Reiss:1982sq,Davidson:1983fy,Chang:1987hz,Berezhiani:1990jj,Berezhiani:1990wn}, e.g.~in theories addressing the observed hierarchical pattern of fermion masses and mixing and the strong CP problem at the same time~\cite{Ema:2016ops,Calibbi:2016hwq}.
In these latter examples, ALP couplings to charged leptons, which are the focus of the present work, are unavoidable, although model-dependent\,\,\,---\,\,\,for more instances of `leptonic' axion and ALP models, see~\cite{Calibbi:2020jvd}. 
As all ALP couplings, these are inversely proportional to the scale $f_a$ at which the symmetry is broken and thus suppressed if this occurs at energies much larger than the lepton mass scale.

ALP interactions with charged leptons can be tested at laboratory experiments either directly or indirectly via the coupling to photons that they unavoidably induce at the loop level. Such laboratory constraints come from measurements of electric~\cite{DiLuzio:2020oah} and magnetic dipole moments\,\,\,---\,\,\,in certain regions of the parameter space, ALP contributions can be large enough as to account for the observed muon $g-2$ discrepancy, see e.g.~\cite{Marciano:2016yhf,Bauer:2019gfk,Buttazzo:2020vfs,Buen-Abad:2021fwq,Liu:2022tqn}\,\,\,---\,\,\,from beam dump experiments~\cite{Essig:2010gu,Dobrich:2015jyk}, and from low-energy $e^+e^-$ colliders such as B factories~\cite{Dolan:2017osp}.\footnote{In presence of flavour-violating couplings, severe constraints are also provided by searches for lepton flavour violation~\cite{Calibbi:2020jvd,Cornella:2019uxs,Bauer:2021mvw,Jho:2022snj}.} 
However, most of these probes tend to lose sensitivity if the ALP is light and weakly-coupled enough to be long-lived, that is, for large values of $f_a$.\footnote{An exception is provided by the search for light dark matter particles in $e^+e^-$ annihilations at the NA64 experiment~\cite{Andreev:2021fzd}, which is also sensitive to an invisible ALP coupling to electrons.} 

The latter situation is best constrained by astrophysical observations. Emission of light particles would, in fact, contribute to the energy loss of stellar systems altering their evolution~\cite{Raffelt:1996wa}. Consequently, very stringent bounds to ALP-electron couplings can be derived from observations of populations of stars such as white dwarfs (WD) and red giants (RG)~\cite{Viaux:2013lha,Bertolami:2014wua}, as long as the ALPs are light enough to be produced inside the star ($m_a \lesssim \ord{10}$~keV) and weakly-coupled enough to avoid to be reabsorbed. 
Hence, these limits do not apply to heavier ALPs and/or ALPs characterised by medium-to-low values of the breaking scale $f_a$, while still long-lived for what concerns laboratory experiments. 
Similarly, the observation of the supernova explosion SN1987A can be used to set limits on ALP couplings to electrons~\cite{Raffelt:1990yz} that are somewhat weaker than those from WD and RG but extend to $m_a \lesssim \ord{100}$~MeV given the much hotter environment. Furthermore, the SN1987A data were recently employed to set bounds to the coupling to muons as well~\cite{Bollig:2020xdr,Croon:2020lrf,Caputo:2021rux}, the latter ones being sensitive also to the low $f_a$ regime. These constraints are however affected by our uncertainty about the supernova explosion mechanism and could even disappear in certain scenarios thereof~\cite{Bar:2019ifz}.

The above discussion shows that it would be desirable to have independent direct laboratory probes of the ALP couplings to leptons, sensitive to the case of a long-lived (and thus invisible) ALP. 
Recent proposals along this direction have focused on the emission of ALPs in meson decays, showing how charged pion~\cite{Altmannshofer:2022izm} and kaon~\cite{Krnjaic:2019rsv} decays can be sensitive to, respectively, the coupling to electrons and muons of a light invisible particle.

The aim of the present note is to show that 
the proposed high-energy $e^+e^-$ accelerators, the Circular Electron-Positron Collider (CEPC)~\cite{CEPCStudyGroup:2018rmc,CEPCStudyGroup:2018ghi} and the $e^+e^-$ stage of the Future Circular Collider (FCC-ee)~\cite{Abada:2019lih,Abada:2019zxq}, can provide a unique opportunity to discover or constrain invisible ALPs through their couplings to charged leptons. Both proposals plan to run for several years at a centre of mass energy around the $Z$ pole, $\sqrt{s}\simeq 91$~GeV, hence operating as ``Tera~Z'' factories, that is, producing more than $10^{12}$ visible $Z$ boson decays at two interaction points.
As the $Z$ boson has a probability of about 10\% of decaying into two charged leptons, the CEPC and FCC-ee would produce more than $10^{11}$ boosted lepton pairs, $\ell^+\ell^-$ ($\ell = e,\,\mu,\,\tau$), thus offering the opportunity to test the emission of an ALP $a$ from the final state leptons in the process $Z\to \ell^+\ell^-a$. In this paper, we focus on the parameter space where $a$ would decay outside the detector hence appearing as missing energy and we simulate the signal and background for $e^+e^- \to Z^{(*)}\to \ell^+\ell^-a$, in order to estimate the sensitivity of CEPC/FCC-ee to the ALP couplings to leptons.

A number of recent works studied possible searches for ALPs at future $e^+e^-$ colliders~\cite{Bauer:2018uxu,Zhang:2021sio,Yue:2022ash,Liu:2022tqn,Lu:2022zbe}, focusing on other ALP-SM interactions and production modes and/or on promptly-decaying ALPs\,\,\,---\,\,\,which is a consequence of the relatively strong couplings necessary e.g.~for an explanation of the muon $g-2$ anomaly. As mentioned above, here we are instead concerned with 
lighter and more weakly-coupled ALPs that would escape the detector unseen.

The rest of the paper is organised as follows. In \sref{sec:model}, we set the notation, review the effective Lagrangian describing ALP-lepton interactions, and discuss the ALP decay modes and lifetime. In \sref{sec:zdecay}, we discuss ALP emission in leptonic $Z$ decays and present an analytical estimate of the number of ALP events to be expected at future $e^+e^-$ colliders. In \sref{sec:result}, we show the result of our simulation and its impact on the ALP parameter space. Finally, in \sref{sec:conclusion}, we summarise and conclude.

\section{ALP couplings to leptons and decays}
\label{sec:model}
In the following, we review the simplified-model approach to lepton-ALP couplings that we employ in our analysis. 
The interactions we are interested in can be written as a dimension-five operator involving a derivative ALP coupling with a leptonic axial current:
\begin{align}
\mathscr{L}_{\text{eff}} \supset \sum_{\ell = e,\mu,\tau} \frac{C_{\ell \ell}^A}{2f_a}\,\partial_\mu a\,(\bar\ell \gamma^\mu\gamma_5\ell)\,,\label{eqn:L}
\end{align}
where $a$ represents the ALP field, $\ell$ denotes charged leptons (that is, $\ell= e,\,\mu,\,\tau$), and $C_{\ell \ell}^A$ are model-dependent dimensionless coefficients that we will be treating as free parameters throughout the paper. The ALP decay constant $f_a$ is a scale related to the spontaneous breaking of the global $U(1)$ symmetry the ALP is associated to. 
Notice that in this work we restrict the discussion to flavour-conserving couplings to leptons. For the very tight constraints that arise in presence of flavour-violating ALP-lepton interactions, in particular from lepton-flavour-violating decays $\ell\to \ell^\prime\, a$ into an invisible ALP, see~\cite{Calibbi:2020jvd,Jho:2022snj}.

Upon integrating by parts and inserting the equations of motion of the lepton fields, the above Lagrangian is equivalent\,\,\,---\,\,\,up to a shift of the anomalous coupling to photons\,\,\,---\,\,\,to the following dimension-four interaction terms involving leptonic pseudoscalar currents, see e.g.~\cite{Bauer:2020jbp}:
\begin{align}
\mathscr{L}_{\text{eff}} \supset - {\rm i} \sum_{\ell} g_{\ell}\, a \,(\bar\ell\gamma_5\ell) \,,\quad 
g_{\ell} \equiv C_{\ell \ell}^A \frac{m_{\ell}}{f_a}\,,
\label{eqn:L2}
\end{align}
which shows that the ALP-lepton couplings are proportional to the lepton mass $m_{\ell}$ and thus larger for heavier generations.

The above lepton-ALP interactions induce, when kinematically allowed, ALP decays into lepton pairs, whose width reads:
\begin{align}
\Gamma(a\to \ell^+ \ell^-)= \frac{m_a}{8\pi}m_{\ell}^2\left(\frac{C_{\ell \ell}^A}{f_a}\right)^2\sqrt{1-\frac{4 m_{\ell}^2}{m_a^2}} 
\,.
\end{align}

The same interactions unavoidably contribute to the effective ALP coupling with photons $E_{\text{eff}}$, which we define as:
\begin{align}
    \mathscr{L}_{\text{eff}} \supset E_{\text{eff}} \frac{\alpha_\text{em}}{4\pi} \frac{a}{f_a} F\widetilde{F}\,.
\end{align}
This quantity depends on a model-dependent UV contribution and, through lepton loops, on the ALP couplings to leptons in \eref{eqn:L}:
\begin{align}
E_{\text{eff}}=E_{\text{UV}}+\sum_{\ell}C_{\ell \ell}^A \,B\left(z\right)\,, \label{eq:Eeff}   
\end{align}
where $z={4m_{\ell}^2}/{m_a^2}$ and the loop function can be written as~\cite{Bauer:2017ris}:
\begin{align}
B(z)=1-zf^2(z)\,,~~~f(z)=\begin{cases}
\arcsin{\frac{1}{\sqrt{z}}},~z\ge 1,\\
\frac{\pi}{2}+\frac{{\rm i}}{2}\ln{\frac{1+\sqrt{1-z}}{1-\sqrt{1-z}}},~z<1\,.
\end{cases}
\end{align}
Notice that the latter contribution correctly decouples in the limit $m_a \ll m_{\ell}$, as $B(z) \to 0$ for $z\to\infty$.
In terms of the above-defined effective coupling, the decay rate into photons reads
\begin{align}
\Gamma(a\to \gamma\gamma)=\frac{\alpha_{\text{em}}^2E_{\text{eff}}^2}{64\pi^3}\frac{m_a^3}{f_a^2}\,.
\end{align}

\begin{figure}[t!]
    \centering
    \includegraphics[width = 0.9\textwidth]{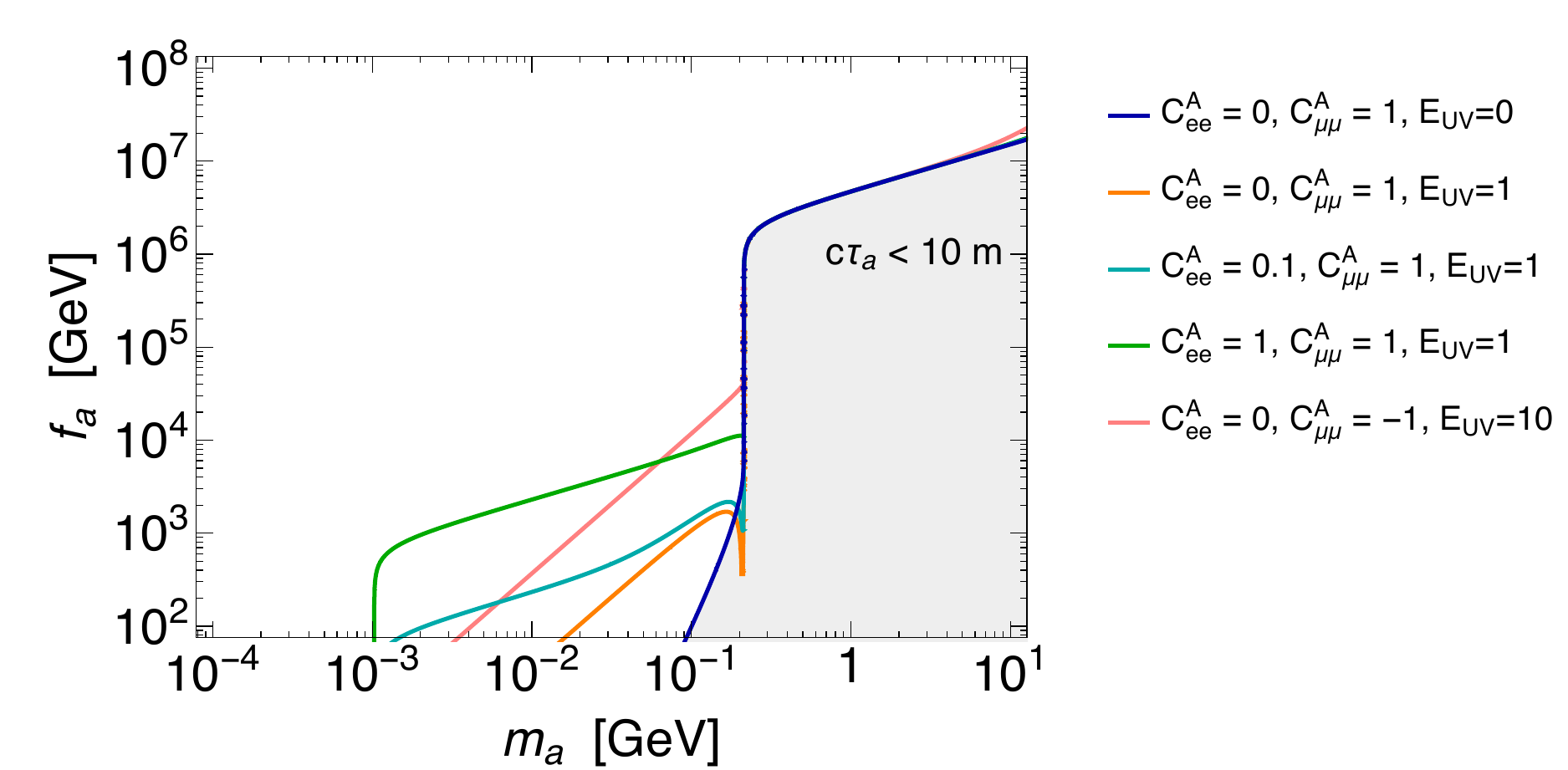}
    \caption{ALP decay length in the ($m_a$,~$f_a$) plane: in the gray region an ALP coupling only to muons mostly decays inside the detector ($c\tau_a < 10$~m). The coloured lines show how the boundary of the region is modified for different choices of the couplings to leptons and photons.}
    \label{fig:lifet}
\end{figure}
Finally, ALP decays into neutrinos $a\to \nu_i\bar\nu_i$ are typically suppressed by the small neutrino masses, hence we set 
\mbox{$\Gamma(a\to \nu_i\bar\nu_i) =0$} throughout the paper.\footnote{Furthermore, ALPs decaying into neutrinos inside a collider detector would still appear as missing energy. Hence, these decay modes would not change the following discussion.} 
Thus we use the following expression for the total decay width of the ALP:
\begin{align}
\Gamma_{a}=\Gamma(a\to \gamma\gamma) + \sum_\ell \theta\left(m_a - 2m_{\ell}\right)\times\Gamma(a\to \ell^+ \ell^-)\,.
\label{eqn:decayALP}
\end{align}

The above expression enables us to assess under which conditions the emitted ALPs preferably decay outside of the detector, thus giving rise to the missing energy signature that we are going to study in the following sections. 
In \fref{fig:lifet}, we highlight in grey the region of the parameter space where the proper decay length $c\tau_a \equiv c/\Gamma_a$ of a `purely muonic' ALP ($C^A_{\mu\mu}=1$, $C^A_{ee}=0$, $E_{\rm UV} =0$)\footnote{Notice that the coupling to tau leptons $C^A_{\tau\tau}$ has no effect on the ALP lifetime as long as $m_a \ll m_\tau$.} is of the order of the size of typical collider experiment detectors or smaller, $c\tau_a < 10$~m. Outside this region, the ALP is typically long-lived. The figure also shows how the boundary of the region varies for different choices of the coupling to electrons $C^A_{ee}$ and the UV-dependent coupling to photons $E_{\rm UV}$. 
As one can see, the ALP is mostly long-lived as long as its mass is below the kinematic threshold of the decay into a muon pair, $m_a < 2\,m_\mu \simeq 210$~MeV. This is also true even in presence of ALP couplings to electrons and photons, at least if they feature a certain hierarchical structure such that $C^A_{ee} \ll C^A_{\mu\mu}$\,\,\,---\,\,\,otherwise the ALP would mostly decay as $a\to e^+e^-$ for low values of the ALP decay constant, $f_a \lesssim 1$~TeV, and $m_a > 2\,m_e$.
For reasons that we will be discussed in the following sections, we will mostly focus on ALPs coupling to muons and set $C^A_{ee}=0$. For definiteness, we will also assume $E_{\text{UV}}=0$, which corresponds to the case of a PNGB of a $U(1)$ symmetry free of electromagnetic anomalies, an example being provided by the majoron, the PNGB associated to the spontaneous breaking of the lepton number~\cite{Chikashige:1980ui,Gelmini:1980re,Georgi:1981pg}. Furthermore, \fref{fig:lifet} shows that $E_{\text{UV}}\neq 0$ would not affect much the region we are interested in where $a$ is long-lived (unless $|E_{\text{UV}}|\gg 1$).
In \sref{sec:result}, we will comment about the impact of relaxing the above assumptions ($C^A_{ee}=0$, $E_{\text{UV}}=0$) on the prospected CEPC/FCC-ee constraints on the ALP parameter space.

\section{ALP emission in leptonic $Z$ decay}
\label{sec:zdecay}

\begin{figure}[t!]
    \centering
    \includegraphics[width =1\textwidth]{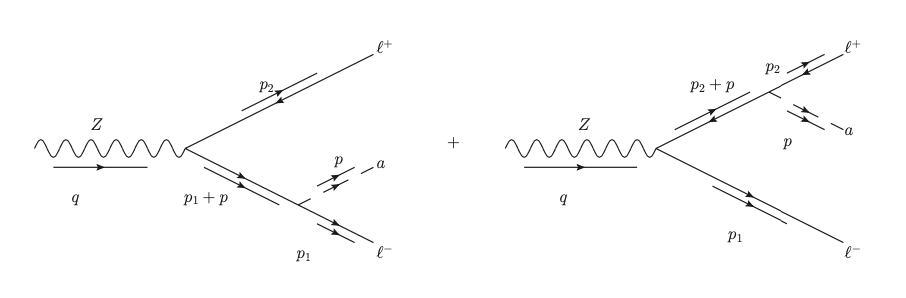}
    \caption{Feynman diagrams of leptonic $Z$ decay with ALP emission.}
    \label{fig:FD}
\end{figure}

As we aim at estimating the sensitivity of CEPC/FCC-ee on ALPs emitted by one of the final state leptons in leptonic $Z$ decays, we present here the analytical expression for the decay width of $Z\to \ell^+\ell^-a$. Following from the diagrams in \fref{fig:FD} and the interaction in the Lagrangian given in \eref{eqn:L}, the partial $Z$ decay width we are interested in reads
\begin{align}
\Gamma(Z\to \ell^+\ell^-a)~= ~\frac{G_F m_{\ell}^2}{192\sqrt{2}\pi^3 m_Z^3}\left(\frac{C_{\ell\ell}^{A}}{f_a}\right)^2
\bigintsss & \dd m_{13}^2\dd m_{23}^2  \bigg[\frac{m_Z^2(m_{23}^4+m_{13}^4)}{m_{23}^2m_{13}^2} \left(2c_{2w}^2-2c_{2w}+1\right) 
\nonumber \\
&  -\left(1-2c_{2w})^2(m_{13}^2+m_{23}^2\right) \bigg],\label{eqn:ZwidthAna}
\end{align}
where $m_Z$ is the $Z$ boson mass, $G_F$ is the Fermi constant, $c_{2w} \equiv \cos 2 \theta_w$ with $\theta_w$ denoting the weak mixing angle, and $m_{13}^2 \equiv (p_1+p)^2$, $m_{23}^2 \equiv (p_2+p)^2$, with the 4-momenta defined as in \fref{fig:FD}. The integration is performed over the usual phase space available to a three-body decay, see e.g.~\cite{10.1093/ptep/ptac097}.

Although the process we are going to simulate is $e^+e^- \to Z^{*} \to \ell^+\ell^- a$, the above analytical expression of the width of on-shell $Z$ decays already allows us to obtain a first approximate assessment of the capability of future colliders. Taking into account the total width of the $Z$ boson, $\Gamma_Z \simeq 2.5$~GeV~\cite{10.1093/ptep/ptac097}, \eref{eqn:ZwidthAna} numerically gives for $m_a \ll m_\ell$:
\begin{align}
\label{eq:Zeea}
    &\text{BR}(Z\to e^+e^-a) ~\approx~  1.2\times 10^{-13}\,\left(\frac{100~\text{GeV}}{f_a/C_{ee}^{A}}\right)^2\,, \\
\label{eq:Zmumua}    
    &\text{BR}(Z\to \mu^+\mu^-a) ~\approx~  2.5\times 10^{-9}\,\left(\frac{100~\text{GeV}}{f_a/C_{\mu\mu}^{A}}\right)^2\,, \\
\label{eq:Ztautaua}        
    &\text{BR}(Z\to \tau^+\tau^-a) ~\approx~  3.3\times 10^{-7}\,\left(\frac{100~\text{GeV}}{f_a/C_{\tau\tau}^{A}}\right)^2\,.
\end{align}
The above values correspond to the largest possible branching ratios that we can arguably expect within viable UV completions of the effective Lagrangian in \eref{eqn:L}, as it may be challenging to formulate such theories for $f_a \lesssim$~100~GeV.

According to the CEPC conceptual design report (CDR)~\cite{CEPCStudyGroup:2018rmc,CEPCStudyGroup:2018ghi},  in the course of the $Z$-pole run, two detectors should collect data amounting to an integrated luminosity of $\mathcal{L} = 50~{\rm ab}^{-1}$, equivalent to approximately $2\times 10^{12}$ visible $Z$ decays. More recent assessments of the expected accelerator performance have updated this figure to $\mathcal{L} = 96~{\rm ab}^{-1}$, that is,  more than $4\times 10^{12}$ $Z$ decays~\cite{CEPCworkshop}. Similarly, the FCC-ee CDR~\cite{Abada:2019lih,Abada:2019zxq} estimates that the collider should deliver in total $\mathcal{L} = 150~{\rm ab}^{-1}$ ($\approx 7\times 10^{12}\,Z$ decays) at two interaction points.
Comparing these planned luminosities with the estimates for the branching ratios in Eqs.~(\ref{eq:Zeea})-(\ref{eq:Ztautaua}), one can see that no signal events are to be expected from $Z\to e^+e^-a$, as a consequence of the process being suppressed by a factor $\sim (m_e/f_a)^2$, see~\eref{eqn:ZwidthAna}. On the other hand, we can foresee that CEPC/FCC-ee would produce up to $\ord{10^4}$ $Z\to \mu^+\mu^-a$ events and $\ord{10^6}$ $Z\to \tau^+\tau^-a$ events. For this reason, in the following, we will mostly focus on $Z\to \mu^+\mu^-a$ and comment about $Z\to \tau^+\tau^-a$ that, despite the larger probability, is much more difficult to disentangle from the background as a consequence of the missing energy due to the neutrinos from tau decays. 

\section{Collider simulation and discussion}
\label{sec:result}

In order to simulate the process we are interested in\,\,\,---\,\,\,$e^+e^- \to  \ell^+\ell^- a$ with a centre of mass energy $\sqrt{s} \simeq 91$~GeV\,\,\,---\,\,\,we implement the Lagrangian in \eref{eqn:L} in \textsc{FeynRules}~\cite{Alloul:2013bka} and we employ the resulting model files~\cite{Degrande:2011ua} within the \textsc{MadGraph} framework~\cite{Alwall:2011uj,Maltoni:2002qb} in order to compute cross sections and generate events for the signal and the relevant SM backgrounds. As a cross check of the model's implementation, we calculated BR($Z\to\ell^+\ell^+a$) with \textsc{MadGraph} finding a very good agreement with the results shown in Eqs.~(\ref{eq:Zeea})-(\ref{eq:Ztautaua}). The simulation we present below makes also use of the parton shower generator \textsc{Pythia}~\cite{Sjostrand:2014zea} and the fast detector simulator \textsc{Delphes}~\cite{deFavereau:2013fsa} adopting the parameters of a typical CEPC detector~\cite{Chen:2017yel}. Finally, we analyse the resulting events by means of \textsc{Root}~\cite{Brun:1997pa} and \textsc{MadAnalysis}~\cite{Conte:2012fm}. 

Having set up the above computation framework, we proceed as follows: (i) We perform a full fast simulation (from event generation to detector response) assuming a given value of $f_a/C^A_{\ell\ell}$ and different values of $m_a$ (although the signal acceptance will turn out to be insensitive to the latter parameter); (ii) Then, we analyse the results of the simulation, in particular the event distributions, to find the suitable kinematical variables to cut on, in order to reduce the background and retain the signal; (iii) Finally, we use the estimated efficiencies of our kinematical cuts to calculate the minimum signal cross section corresponding to a detectable number of events and obtain from it to expected limits on the model's parameters.

\begin{table}[t]
    \centering
    \renewcommand{\arraystretch}{1.3}
    \begin{tabular}{c c c c} \hline\hline
          &  $\sigma(e^+e^-\to  \ell^+\ell^-a)$ & $\sigma(e^+e^-\to  \ell^+\ell^-\nu\bar{\nu})$ & $\sigma(e^+e^-\to  \ell^+\ell^-(\gamma))$\\ \hline
          $ \ell = e$ &$7.1\times 10^{-9}$~pb &  $1.9 \times 10^{-3}$~pb & 4490~pb\\ 
       $ \ell = \mu$ &$7.6\times 10^{-5}$~pb & $2.9 \times 10^{-4}$~pb & 2024~pb \\ 
        $ \ell = \tau$ &$1.1\times 10^{-2}$~pb & $2.8 \times 10^{-4}$~pb & 2020~pb\\ \hline\hline
    \end{tabular}
    \caption{LO cross sections of signal (for $f_a/C^A_{\ell\ell}=100$~GeV, $m_a=10^{-6}$~GeV) and background processes for $e^+e^-$ collisions with $\sqrt{s}=91.2$~GeV, cf.~the main text for details.}
    \label{tab:cross section}
\end{table}   

The ALP appears as missing energy inside the detector, since we focus on the parameter space where it is characterised by a long lifetime, as shown in \sref{sec:model}. Therefore, the main physical background for our processes is given by $e^+e^- \to \ell^+ \ell^-\nu\bar\nu$, typically stemming from the rare four-body decay $Z\to \ell^+ \ell^-\nu\bar\nu$. The cross sections of signal (for representative choices of the ALP parameters) and background processes, as calculated at leading order (LO) by \textsc{Madgraph}, are displayed in \tref{tab:cross section}. 
In addition, there is a sizeable probability of radiation from the initial or final state leptons of one or more photons, $e^+e^- \to \ell^+ \ell^-\gamma$. These processes can fake our signal if the photons, especially the soft ones, get undetected. In order to account for such an instrumental background, we generate large samples of $e^+e^-\to \ell^+\ell^-$ events and let \textsc{Pythia} and \textsc{Delphes} simulate, respectively, the electromagnetic showers and the detector response. As a reference, the LO cross sections for $e^+e^-\to \ell^+\ell^-$ are also shown in \tref{tab:cross section}. 
Finally, notice that the process $e^+e^-\to \tau^+\tau^-$ followed by $\tau^\pm\to\mu^\pm \nu\bar\nu$ also provides a potentially important background for the search for $e^+e^- \to  \mu^+\mu^- a$. From \tref{tab:cross section}, we see that, considering that $\text{BR}(\tau^\pm\to\mu^\pm \nu\bar\nu) \simeq 17.4\%$~\cite{ParticleDataGroup:2022pth}, the LO cross section of such background before cuts is $\approx 61$~pb.

\begin{figure}[t!]
    \centering
    \subfigure[MET in $e^+e^-\to Z^*\to \mu^+ \mu^-+X$]{\includegraphics[width=0.45\textwidth]{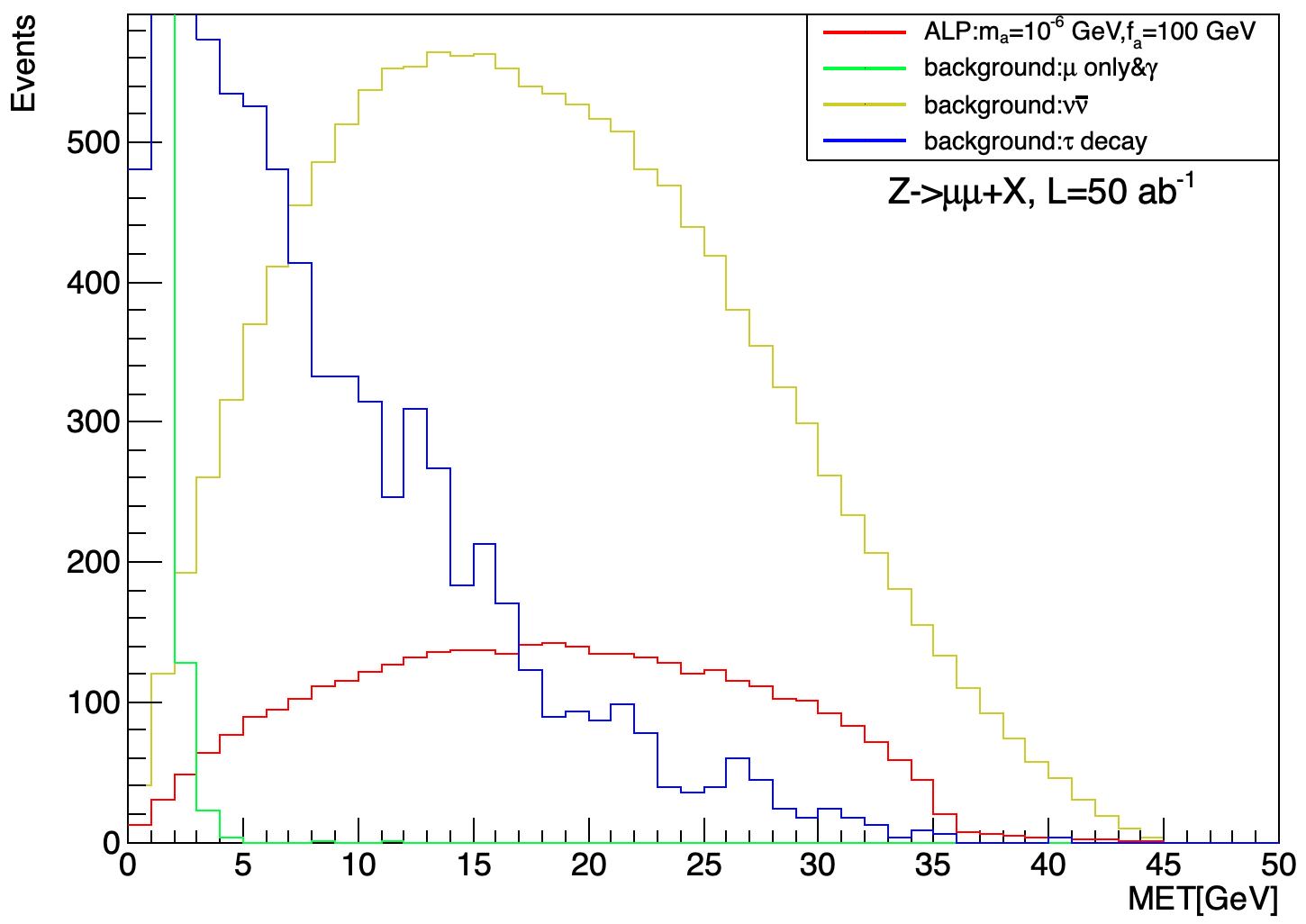}\label{fig:metplot1}}
    \hfill
    \subfigure[PT in $e^+e^-\to Z^*\to \mu^+ \mu^-+X$]{\includegraphics[width=0.45\textwidth]{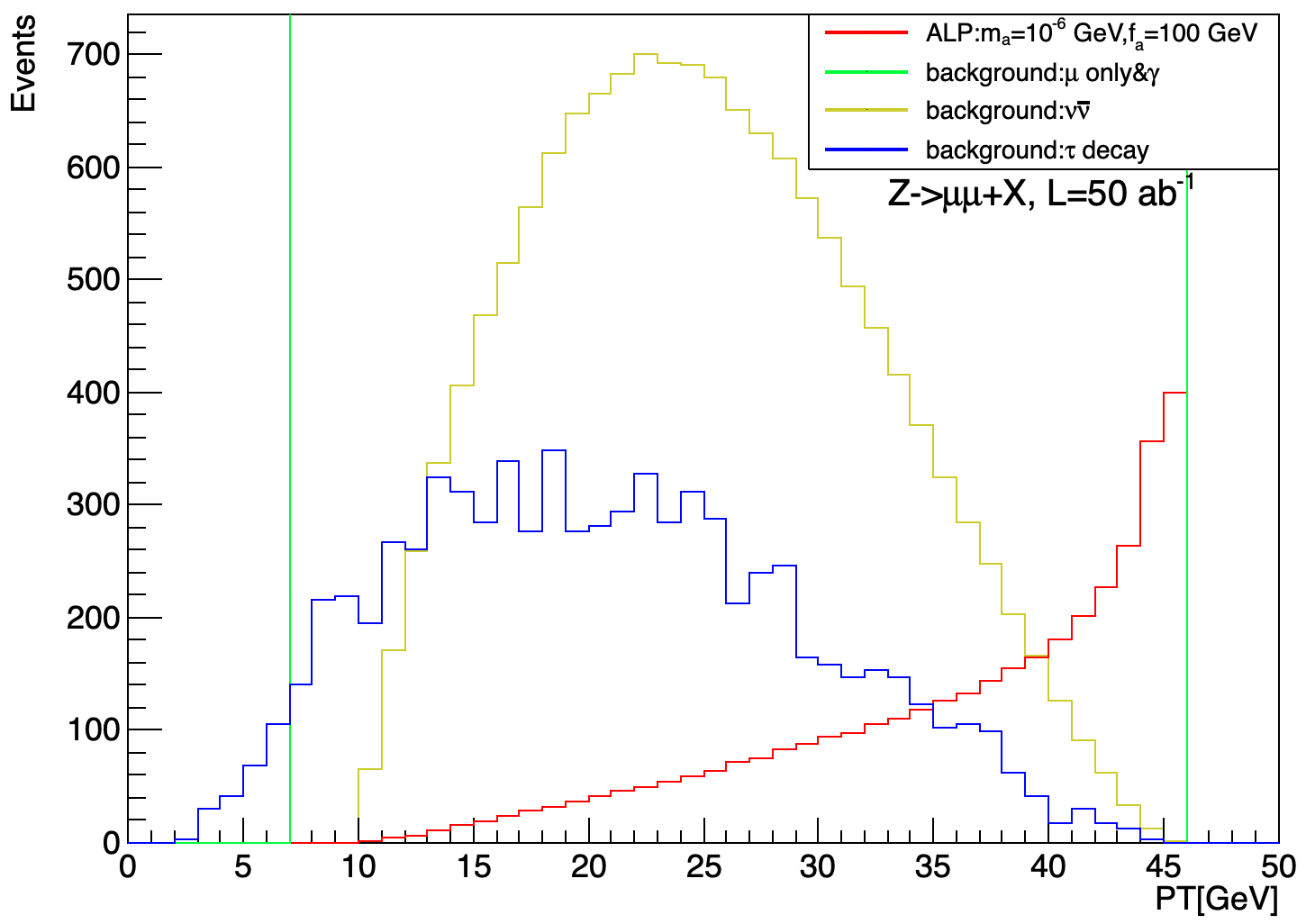}\label{fig:ptplot1}}
    \subfigure[MET in $e^+e^-\to Z^*\to \tau^+ \tau^-+X$]{\includegraphics[width=0.45\textwidth]{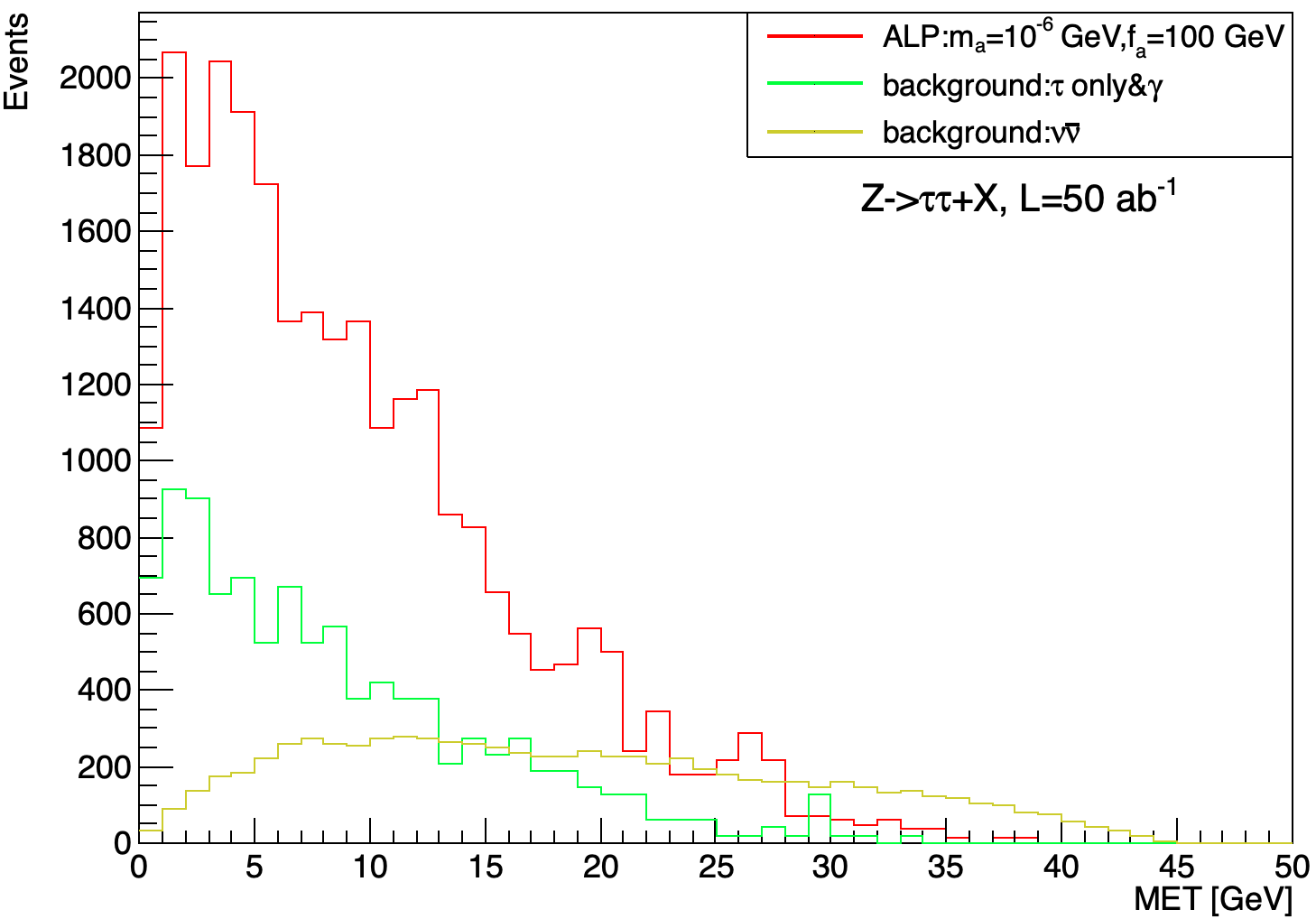}\label{fig:metplot2}}
    \hfill
    \subfigure[PT in $e^+e^-\to Z^*\to \tau^+ \tau^-+X$]{\includegraphics[width=0.45\textwidth]{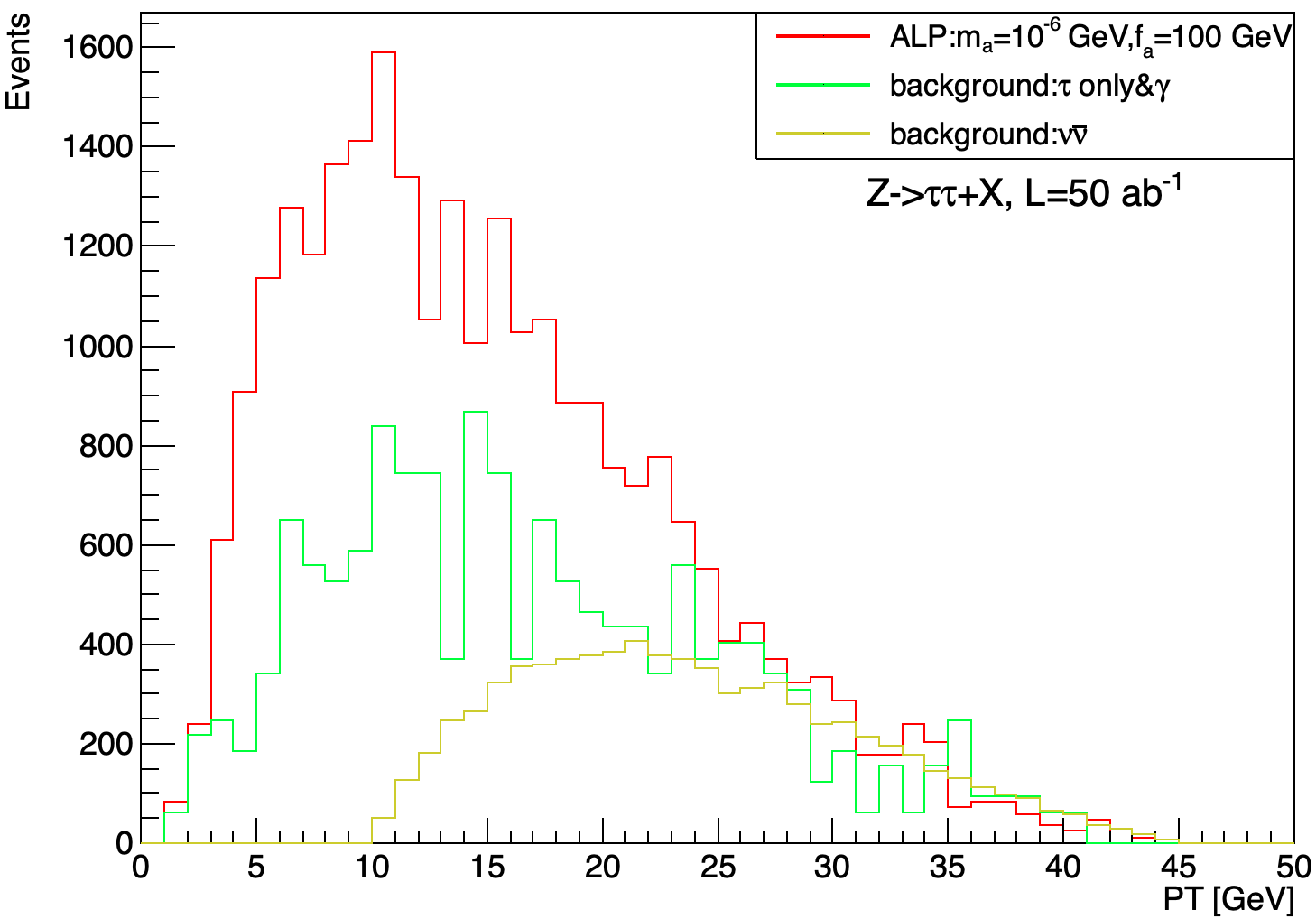}\label{fig:ptplot2}}
    \caption{Distribution, after the basic selection cuts, of missing transverse energy (MET) and transverse momentum of the most energetic lepton (PT) for $e^+e^-\to \ell^+ \ell^-a$ (signal), $e^+e^-\to \ell^+ \ell^-\nu\bar\nu$ and $e^+e^-\to\ell^+\ell^-(\gamma)$ (background), with $\ell = \mu$ (first row), $\ell = \tau$ with $\tau^+\tau^-$ decaying leptonically to different flavour leptons (second row).
    Note that (unlike $e^+e^-\to \ell^+ \ell^-a$ and $e^+e^-\to \ell^+ \ell^-\nu\bar\nu$) the $e^+e^-\to \ell^+ \ell^-$ processes are not normalised according to their cross section, otherwise the large rates would not let the distribution be clearly displayed.}
    \label{fig:metplot}
\end{figure}

The cross section of $e^+e^-\to  e^+e^-a$ displayed in \tref{tab:cross section} confirms that, 
as argued by the analytical estimates in \sref{sec:zdecay},  constraining ALP-electron interactions is beyond the sensitivity of Tera~Z factories.
Hence we only need to simulate the processes $e^+e^-\to \mu^+\mu^-a$ and $e^+e^- \to \tau^+\tau^-a$.
\tref{tab:cross section} shows that the cross section of the latter process is relatively sizeable, in fact larger than the SM process $\sigma(e^+e^-\to \tau^+\tau^-\nu\bar\nu)$ as long as $f_a \lesssim 600$~GeV, as follows from the scaling $\sigma(e^+e^-\to \tau^+\tau^-a) \sim (m_\tau/f_a)^2$.
The challenge of this mode is rather related to the large amount of missing energy coming from the neutrinos in the tau decays that makes it difficult to detect the missing energy associated to possible ALP emission. This is illustrated in the second row of \fref{fig:metplot}, where the distribution of missing transverse energy (MET) and transverse momentum of the most energetic lepton (PT) are shown for events with taus decaying leptonically into different flavour leptons, $\tau^+\tau^- \to e^+\mu^- +$~MET or $e^-\mu^+ +$~MET. As we can see, both the MET and PT distributions from $e^+e^-\to \tau^+\tau^-a$ and $e^+e^-\to \tau^+\tau^-$ are similar. Therefore, in order to search for an ALP signal, one should rather rely on hadronically decaying taus, which requires modelling the tau tagging capabilities of the future detectors and the related uncertainty. We postpone such an analysis to future work and focus in the following on the muon mode only.

We select $\mu^+\mu^- +$~MET events imposing the following basic angular and isolation requirements: $|\eta_{\mu_{i}}|<2.5$ for the pseudorapidity of both muons, and $\theta_{\mu_1\mu_2}>0.2$ for the separation angle between them. We label as $\mu_1$ ($\mu_2$) the most (least) energetic particle of the $\mu^+\mu^-$ pair.
The first row of \fref{fig:metplot} clearly shows that the MET distribution provides an handle to disentangle the signal events $\mu^+\mu^-a$ from the background stemming from $e^+e^- \to \mu^+\mu^-(\gamma)$ (where the photons are missed), while the PT distribution can be used to tame the background from $e^+e^- \to \mu^+\mu^- \nu\bar\nu$ and $e^+e^- \to \tau^+\tau^- \to \mu^+\mu^- 4\nu$.
In particular, a cut MET~$>10$~GeV along with the following PT cut can  eliminate $e^+e^-\to \mu^+\mu^-(\gamma)$ with undetected photons entirely.
On the other hand, an upper cut MET~$<28$~GeV can help reduce the $\mu^+\mu^-\nu\bar\nu$ background retaining most of the signal events, cf.~\fref{fig:metplot1}.
Secondly, a cut on the momentum of the most energetic muon $\mu_1$, PT~$>$~43~GeV, enables us to efficiently distinguish the signal from the neutrino and tau background, as shown in~\fref{fig:ptplot1}.
To summarise, we impose the following requirements on our simulated events:
\begin{itemize}
    \item {\bf Basic selection}: $\theta_{\mu_1\mu_2}>$~0.2, $|\eta_{\mu_i}|<2.5$;
    \item {\bf Cut 1}: missing energy, 10~GeV~$<$~MET~$<$~28~GeV; 
    \item {\bf Cut 2}: $\mu_1$ momentum, PT ~$>$~43~GeV.
\end{itemize}
%
\begin{table}[t!]
    \centering
    \renewcommand{\arraystretch}{1.3}
    \begin{tabular}{ccccc}     \hline\hline
        \multicolumn{2}{c}{} & Basic cuts & Cut~1 & Cut~2\\  \hline
        \multirow{2}*{ ALP Signal}& acceptance & 0.88 & 0.56 & 0.14\\ 
        ~ & $\sigma$ [$\times 10^{-5}$~pb] & 6.7 & 4.3 & 1.1 \\ \hline
        \multirow{2}*{$\nu \bar{\nu}$ background}& acceptance & 0.91 & 0.57 & 0.0015 \\ 
        ~ & $\sigma$ [$\times 10^{-4}$~pb] & 2.6 & 1.7 & 0.0044 \\ \hline
        \multirow{2}*{$\tau$ decay background}& acceptance & 0.025 & 0.008 & 0 \\ 
        ~ & $\sigma$ [pb] & 51.0 & 16.8 & 0 \\ \hline
        {$\gamma$ background}& acceptance & 0.94 & 2.5$\times 10^{-6}$ & 0\\ \hline\hline
    \end{tabular}
    \caption{Signal and background cross section and acceptance after the cuts described in the main text. The signal cross section was calculated assuming $f_a/C^A_{\mu\mu}=100$~GeV.
    }
    \label{tab:result_cut}
\end{table}

The efficiency of the above cuts is displayed in \tref{tab:result_cut}. As we can see, they can effectively remove the background while retaining about 14\% of the signal events.
We can use the resulting acceptance of signal and background to estimate the significance of the proposed search for a given value of the integrated luminosity $\mathcal{L}$.
As customary, we define the effective significance of the signal in terms of the number of signal ($n_s$) and background ($n_b$) events after the cuts as
\begin{eqnarray}
s=\frac{n_s}{\sqrt{n_b+n_s}}\,.
\end{eqnarray}
Setting $s=2$ in this expression, we can derive the 95\% confidence level~(CL) upper limit on the number of signal events that, for a given $\mathcal{L}$, the future Tera~Z factories can set. Similarly, setting $s=5$, we obtain the value of $n_s$ corresponding to a $5\sigma$ discovery. This information can be translated into limits on the signal cross section (and equivalently on the decay rate of $Z\to \mu^+\mu^-a$) and consequently on the ALP coupling to muons. 
In \tref{tab:result_fin}, we show the resulting limits for three benchmark values for the integrated luminosity that CEPC and FCC-ee expect to collect at two interaction points from the operation of the $Z$-pole run, as reported in~\cite{CEPCStudyGroup:2018rmc,CEPCStudyGroup:2018ghi,CEPCworkshop,Abada:2019lih,Abada:2019zxq}.
In particular, the table displays the 95\% CL limits we obtain on the signal cross section ($\equiv\sigma_{95}$), on $f_a /C^A_{\mu\mu}$ ($\equiv f_{95}$), and on $\text{BR}(Z\to \mu^+\mu^-a)$ ($\equiv$\,BR$_{95}$), as well as the corresponding $5\sigma$ discovery values ($\sigma_{5\sigma}$, $f_{5\sigma}$, BR$_{5\sigma}$).
As we can see, a search of this kind at a Tera~Z factory can be sensitive to ALP decay constants up to about 900~GeV and yield $\text{BR}(Z\to \mu^+\mu^-a)\lesssim 3\times10^{-11}$ as the upper bound on the branching ratio for the $Z$ decay into a muonic ALP. 

\begin{table}[t!]
    \centering
    \renewcommand{\arraystretch}{1.3}
    \begin{tabular}{lccc}
    \hline\hline
     & $\mathcal{L} =50\,\text{ab}^{-1}$ & $\mathcal{L} =100\,\text{ab}^{-1}$ & $\mathcal{L} =150\,\text{ab}^{-1}$  \\ \hline
    $\sigma_{95}$  ($\sigma_{5\sigma}$) [$\times 10^{-6}$\,pb] & 1.7 (5.6) & 1.1 (3.4) & 0.87 (2.6)\\ 
    $f_{95}$ ($f_{5\sigma}$) [GeV]& 680 (370) & 834 (473) & 936 (541)  \\ 
        $\text{BR}_{95}$ ($\text{BR}_{5\sigma}$) [$\times 10^{-11}$]& 5.4 (18) & 3.6 (11) & 2.8 (8.5) \\ \hline\hline
    \end{tabular}
    \caption{Projected $95\%$ CL limit ($5\sigma$-discovery sensitivity) on the signal cross section, the ALP-muon coupling $f_a/C^A_{\mu\mu}$, and $\text{BR}(Z\to \mu^+\mu^-\,a)$. See the text for details.}
    \label{tab:result_fin}
\end{table}

\begin{figure}[t!]
    \centering
    \includegraphics[width = 0.75\textwidth]{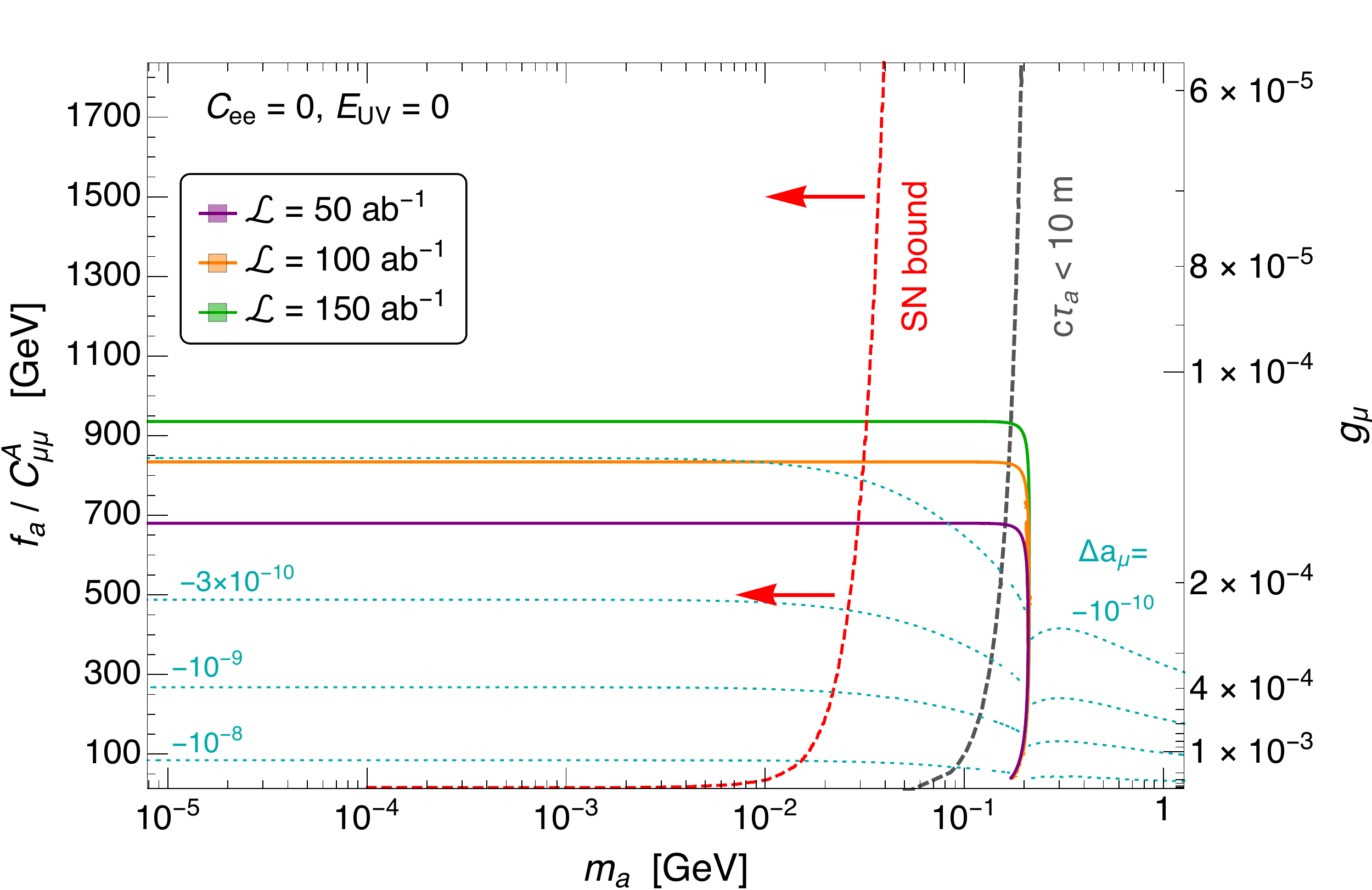}
    \caption{Prospected CEPC/FCC-ee 95\% CL exclusion on the $(m_a,\,f_a/C^A_{\mu\mu})$ plane for a muonic ALP ($C^A_{ee}=0$, $E_\text{UV}=0$) with different assumptions for the integrated luminosity $\mathcal{L}$, as indicated. On the right side of the dashed grey line the proper decay length of the ALP, as calculated using \eref{eqn:decayALP}, is $c\tau_a <10$~m. The region to the left of the red dashed line is excluded by SN1987A data, according to the analysis in \cite{Croon:2020lrf}. The dotted cyan contours show the ALP contribution to the anomalous magnetic moment of the muon, $\Delta a_\mu \equiv (g-2)_\mu/2$. See the main text for details. }
    \label{fig:final}
\end{figure}

The results of \tref{tab:result_fin} can be employed to show the sensitivity reach of CEPC/FCC-ee on the ALP parameter space.
In \fref{fig:final}, we plot the estimated 95\% CL exclusion lines on the $(m_a,\,f_a/C^A_{\mu\mu})$ plane for the case of an anomaly-free ALP ($E_{\rm UV}=0$) whose coupling to electron vanishes ($C^A_{ee}=0$), showing three representative values of $\mathcal{L}$ as in \tref{tab:result_fin}. 
Notice that the limits in the figure can be also read in terms of the dimensionless coupling $g_\mu$, as defined in \eref{eqn:L2}, that can be excluded down to $g_\mu \approx 10^{-4}$.
In \fref{fig:final}, the number of signal events have been convoluted with the probability that the ALP decays inside a typical detector (for definiteness, we assumed a flight distance $L=10$~m), taking into account the proper decay length $c\tau_a$ calculated from \eref{eqn:decayALP}. As expected from the discussion in \sref{sec:model}, this cuts off the sensitivity of our search for an invisible ALP around the kinematical threshold of the decay $a\to \mu^+\mu^-$. For $m_a \gtrsim 2\,m_\mu$, the model can be tested through searches for $Z \to \mu^+\mu^- \mu^+\mu^-$\,\,\,---\,\,\,with the invariant mass $m^2_{\mu\mu}$ of one of the muon pairs featuring a resonance in correspondence of $m^2_{\mu\mu}=m_a^2$\,\,\,---\,\,\,as discussed in \cite{Liu:2022tqn}.
We note that \fref{fig:lifet} shows how the parameter space that our search is sensitive to would be reduced if we switch on $C^A_{ee}$ and $E_{\rm UV}$. In particular, for a coupling to electrons of the same order as the one to muons,  $C^A_{ee}\approx C^A_{\mu\mu}$, our search would lose sensitivity for $m_a \gtrsim 2\,m_e \approx 1$~MeV, where ALPs could be sought through the search for a di-electron resonance in $Z \to \mu^+\mu^- e^+e^-$. However, the sensitivity loss could be evaded for short-lived ALPs with mass $m_a>2\,m_e$ or even $m_a> 2\,m_\mu$, if they mostly decay into invisible particles, for instance belonging to the dark matter sector\,\,---\,\,see e.g.~the model in \cite{Buttazzo:2020vfs}. In such a case, the search we propose would be sensitive to heavier ALPs as well.

\begin{figure}[t!]
    \centering
    \includegraphics[width = 0.75\textwidth]{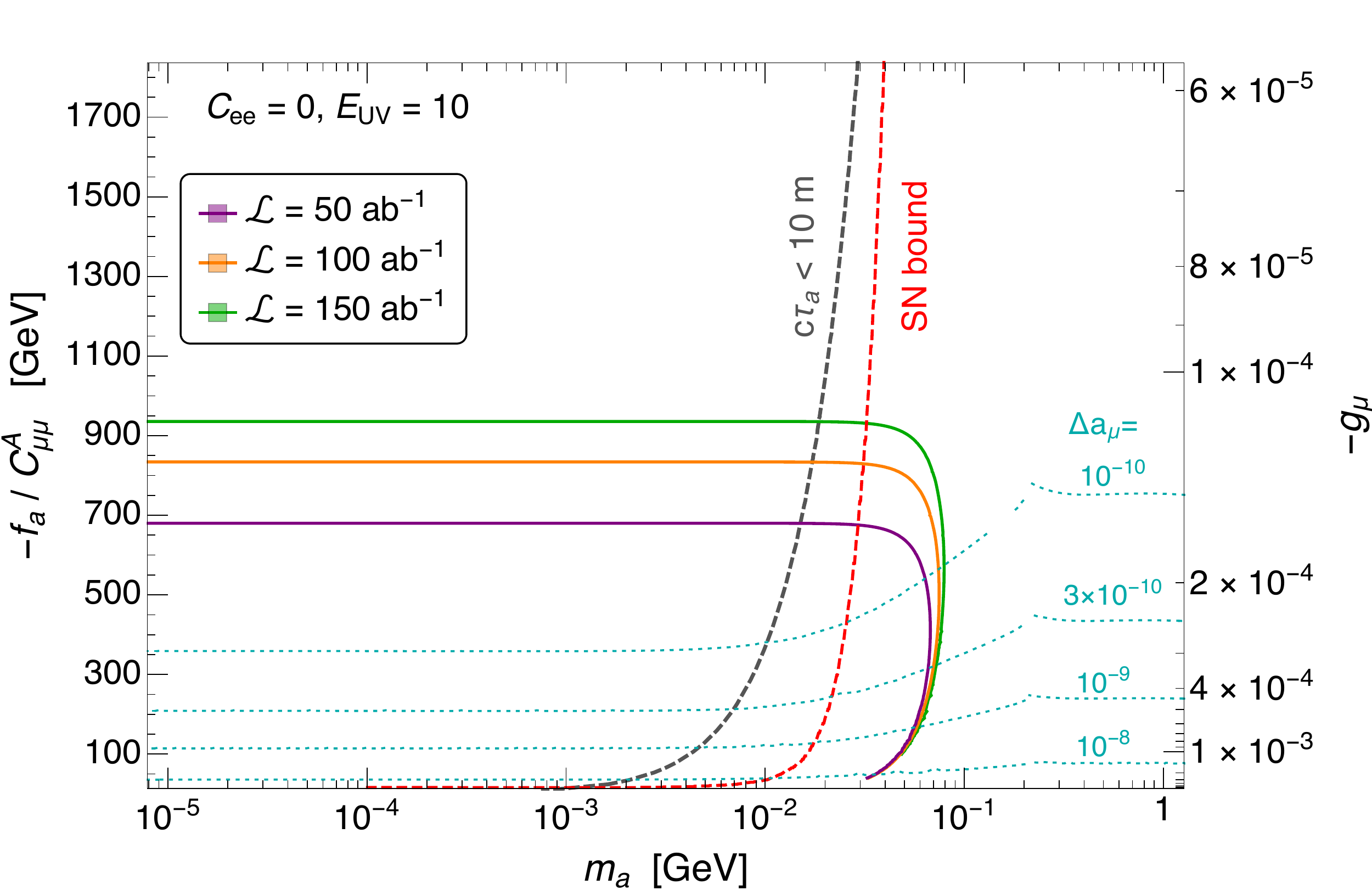}
    \caption{Same as \fref{fig:final} for an ALP with large coupling to photons ($E_\text{UV}=10$) and $C^A_{\mu\mu}<0$, a choice that can yield a potentially large positive contribution to $\Delta a_\mu$.}
    \label{fig:final2}
\end{figure}

As a reference, in \fref{fig:final}, we also display contours of $\Delta a_\mu$, that is, the ALP correction to the anomalous magnetic moment of the muon $a_\mu \equiv (g-2)_\mu/2$, as calculated using the formulae for the 1-loop and 2-loop contributions in \cite{Buen-Abad:2021fwq,Liu:2022tqn}\,\,---\,\,see also \cite{Marciano:2016yhf,Bauer:2019gfk,Buttazzo:2020vfs}. As is well known, without a substantial coupling to photons, the ALP-muon interaction gives raise to a negative $\Delta a_\mu$. 
However, one can see that our search for $e^+e^- \to \mu^+\mu^-a$ could test the parameter space for $|\Delta a_\mu| \gtrsim 10^{-10}$, that is, up to values that do not constitute a substantial shift from the SM because well within the current experimental and theoretical uncertainties, which are both $\approx (4-5)\times 10^{-10}$.
In fact, considering the presently inconclusive status of the SM prediction for what concerns the leading hadronic contribution~\cite{Aoyama:2020ynm,Borsanyi:2020mff}, we can set the following conservative 95\% CL bound solely based on the experimental average~\cite{fnal}: $|\Delta a_\mu|  < 8.2\times 10^{-10}$. As we can see from \fref{fig:final}, this disfavours the parameter space below approximately $f_a/C^A_{\mu\mu} \approx 300$~GeV. Even the tighter bound $ -3.4\times 10^{-10}<\Delta a_\mu  < 2.5\times 10^{-9}$ based on the lattice prediction in~\cite{Borsanyi:2020mff}\,\,---\,\,which is 1.5$\sigma$ below the experimental measurement\,\,---\,\,results in a limit ($f_a/C^A_{\mu\mu} \gtrsim 500$~GeV) weaker than our estimated CEPC/FCC-ee prospect. Taking instead the prediction for the hadronic vacuum polarisation based on dispersive relations~\cite{Aoyama:2020ynm}, new physics inducing a negative $\Delta a_\mu$ is strongly disfavoured, but so is the Standard Model itself. If this will prove to be the case, a positive ALP contribution to the $a_\mu$ can be achieved with a large coupling to photons with sign opposite to that to muons~\cite{Marciano:2016yhf,Bauer:2019gfk,Buttazzo:2020vfs,Buen-Abad:2021fwq,Liu:2022tqn}. An example is provided in \fref{fig:final2}, where we take $E_\text{UV} = 10$ in \eref{eq:Eeff}
and $C^A_{\mu\mu} <0 $. As we can see, a contribution within the $2\sigma$ region favoured by \cite{Aoyama:2020ynm}, $ 1.3\times 10^{-9}<\Delta a_\mu  < 3.7\times 10^{-9}$, would require $|f_A /C^A_{\mu\mu}| = \mathcal{O}(100)$~GeV. This regime can be easily tested by our search as long as $m_a \lesssim 0.1$~GeV, while heavier ALPs would decay promptly to photons unless, again, a large coupling to dark sector particles let them decay mostly invisibly, in which case our search sensitivity would extend to larger values of $m_a$.

Finally, Figures \ref{fig:final} and \ref{fig:final2} also shows the SN1987A constraint on light particles coupling to muons, as calculated in~\cite{Croon:2020lrf}, which excludes the displayed parameter space for \mbox{$f_a/C^A_{\mu\mu}\gtrsim 50$~GeV}, \mbox{$m_a \lesssim \mathcal{O}(10)$~MeV}. For models with $C^A_{ee}\neq 0$, supernova constraints and other star cooling bounds associated to ALP emission from electrons instead lose sensitivity for $f_a/C^A_{ee}\lesssim 10^5$~GeV (see e.g.~\cite{Calibbi:2020jvd}) leaving the range of parameters accessible at colliders unconstrained.
This observation, as well as the interplay between our exclusion lines and the SN1987A bound in Figures \ref{fig:final} and \ref{fig:final2}, shows how searches for long-lived ALPs in $e^+e^- \to Z^{(*)}\to \ell^+\ell^-a$ at future colliders may be nicely complementary to astrophysical probes of such light particles, in particular in view of the uncertainty of the supernova constraint~\cite{Bar:2019ifz}.\footnote{See also~\cite{Caputo:2021rux} for bounds derived from the supernova explosion energy that are relevant in the trapping regime and independent of the explosion mechanism.}

\section{Summary and conclusions}
\label{sec:conclusion}

Axion-like particles are common by-products of new physics models addressing outstanding problems of the Standard Model. In this note, motivated by the fact that interactions with SM leptons are unavoidable for a wide range of well-motivated  ALP scenarios, we have discussed the capability of the proposed high-energy $e^+e^-$ colliders CEPC and FCC-ee of probing this class of particles through their couplings with charged leptons. 
We considered as production process the ALP emission from  leptons in the large samples of $Z\to \ell^+\ell^-$ decays that  CEPC and FCC-ee will collect operating as Tera~Z factories, focusing in particular on the case of long-lived ALPs that would appear as missing energy in the detector.

We first calculated the rate of $Z\to \ell^+\ell^-a$, in order to analytically estimate the number of events that one can expect at Tera~Z factories. 
Then we employed a fast simulation framework (\textsc{Madgraph}+\textsc{Pythia}+\textsc{Delphes}) to compute the LO cross section of $e^+ e^-  \to \ell^+\ell^-a$ and that of the relevant backgrounds, as well as to generate the corresponding events. The main findings of our work can be summarised as follows.
\begin{itemize}
    \item As is already clear from the analytical estimates in \sref{sec:zdecay}, the number of $e^+ e^-  \to e^+e^-a$ events that a Tera~Z can yield is too low to constrain the ALP coupling to electrons, unless $f_a/C^A_{ee}\ll$\,100~GeV\,\,\,---\,\,\,see \eref{eq:Zeea}\,\,\,---\,\,\,a situation that may be difficult to map to a viable UV-complete model.
    \item Concerning the muon mode, the signal $e^+ e^-  \to \mu^+\mu^-a$ can emerge from the background constituted by $\tau^+\tau^-$, $\mu^+\mu^-\nu\bar\nu$, and $\mu^+\mu^-\gamma$ (where $\gamma$ gets undetected), if one applies simple cuts on the missing energy of the event and the momentum of the most energetic muon, see \tref{tab:result_cut}.
    \item We estimate that such a crude ``cut and count'' search\,\,\,---\,\,\,that could be certainly improved and optimised\,\,\,---\,\,\,may be sensitive to ALP-muon interactions characterised by $f_a/C^A_{\mu\mu} \lesssim$\,1~TeV, or equivalently $g_\mu \gtrsim 10^{-4}$, corresponding to $\text{BR}(Z \to \mu^+\mu^-a) \gtrsim 3\times 10^{-11}$\,\,\,---\,\,\,see \tref{tab:result_fin} and \fref{fig:final}. Our summary plot  in~\fref{fig:final} also shows that, unless the ALP mostly decays into dark sector particles, our search would lose sensitivity for $m_a \gtrsim 2\,m_\mu$\,\,\,---\,\,\,where Tera~Z factories may instead search for $Z \to \mu^+\mu^-a \to \mu^+\mu^-\mu^+\mu^-$\,\,\,---\,\,\,and highlights the interplay between collider searches and astrophysical constraints (from SN1987A).
    \item The tau channel $e^+ e^-  \to \tau^+\tau^- a$ enjoys a two order of magnitude enhancement compared to the muon one, $\sim(m_\tau/m_\mu)^2$, such that $\sigma(e^+ e^-  \to \tau^+\tau^-a) > \sigma(e^+ e^-  \to \tau^+\tau^-\nu\bar\nu)$ for $f_a/C^A_{\tau\tau} \lesssim$\,600~GeV. Nevertheless, the missing energy associated to the ALP is way more difficult to detect due to that carried by neutrinos from $\tau^+\tau^-$ decays. As a consequence, a dedicated study focusing on hadronic taus and the expected tagging efficiencies of the CEPC/FCC-ee detectors would be necessary to assess the Tera~Z sensitivity on $f_a/C^A_{\tau\tau}$.
\end{itemize}

In conclusion, we believe that the above results reinforce the physics case of a $Z$-pole run of the proposed $e^+e^-$ colliders. Indeed, our study provides a further example of how Tera Z factories, besides performing very precise electrowek measurements and conducting an excellent particle physics programme (see~\cite{CEPCPhysicsStudyGroup:2022uwl} for a recent overview), could also effectively contribute to searches for new physics particles.

\section*{Acknowledgements}
We would like thank Prof.~Chen Ning for valuable discussions and Wang Bin for technical support.
This work was supported by the Innovation and Entrepreneurship Training Program for College Students of Tianjin (No.~202110055325). LC is partially supported by the National Natural Science Foundation of China under the Grant No.~12035008 and No.~12211530479.


\bibliographystyle{JHEP}
\bibliography{refs}

\end{document}